\documentclass[11pt]{article}
\usepackage{geometry}
\usepackage{amsmath, amssymb}
\usepackage[retainorgcmds]{IEEEtrantools}
\usepackage[all,knot,poly]{xy}
\usepackage{rotating}

\usepackage{graphicx}	
\usepackage{float}
\usepackage[utf8]{inputenc}
\usepackage{multirow}
\usepackage{array}
\usepackage{mathtools}
\usepackage{makecell}
\usepackage{diagbox}
\usepackage{soul}
\usepackage{color}
\numberwithin{equation}{section}
\usepackage{multicol,caption}
\usepackage{empheq}
\usepackage{enumitem}
\usepackage{verbatim}
\usepackage{microtype}
\usepackage[super,sort&compress]{natbib} %puts citations in superscript (does not support brackets)

\title{A Possible Relation Between the Magnetic Pole and the Gravitational Field}
\author{Robert J. Finkelstein}
\date{\emph{Physics \& Astronomy, University of California, Los Angeles, \\
475 Portola Plaza, Los Angeles, California 90095, USA \\
finkel@physics.ucla.edu}}

\begin{document}

\setlength{\baselineskip}{1.6\baselineskip}

\maketitle

\begin{abstract}
\setlength{\baselineskip}{1.6\baselineskip}

We reformulate the quantization of the gravitational field and its sources, including the electric and magnetic fields as they appear in the knot algebra.
\end{abstract}

\section{Introduction}
\paragraph{}We have proposed to augment the standard model by a ``knot algebra" that will define new ``elementary particles".

Dirac \cite{Dirac1} and Schwinger \cite{Schwinger1} have both proposed adding hypothetical magnetic poles to the Quantized Maxwell Theory. We have shown how magnetic poles might be expected already in the context of the ``Knot Algebra", where the field operators of the standard model acquire the factors $D^j_{mm'}(q,z)$, which are the irreducible representations of the quantum group $SL_q(2)$. This extension of the standard model is topological and could unlock magnetic poles. At the same time, because of advances in observational astronomy, the quantization of the gravitational field itself has become a topological challenge, ie, in most of the already explored universe, the influence of the gravitational field has been believed until recently to be too weak to be topologically significant. There is now, however, a growing opinion that different topologies may describe the consequences of earlier and major astronomical events, including gravitational collapse and nuclear explosions, as illustrated by the Gamow ``Big Bang" \cite{Hawking}. With the recent discovery of gravitational waves, physical spacetime may have become a field of speculation about its topological structure.

Some of these topological possibilities are discussed by van der Bij\cite{Bij1} in his review of recent experimental results at CERN.  It is also there pointed out that physical spacetime at earlier times may have been 3-dimensional\cite{carlip} before entering its present phase of physical expansion. Both the new topologies and the new possibilities offered by a hypothetical earlier 2-d space were discussed by us in earlier works describing the knot model\cite{Fink1}.

\section{Knot Model \cite{Fink1} \cite{finkelstein14a} }
\paragraph{} The original purpose of the knot algebra was to provide additional degrees of freedom to the Standard Model. Since the Knot Model associates a classical knot (N, w, r) with every quantum ``knot state" $(j, m, m')$ of the standard model, it is necessary to add an odd number, o, to a defining equation of the ``knot algebra" as follows:
\begin{equation}
    \boxed{(j, m, m')_q = \frac{1}{2}(N, w, r+o)}
    \label{knotAssociation}
\end{equation}
where w and r are of opposite parity while $m$ and $m'$ are of the same parity. Here q is the deformation parameter of $SL_q(2)$

We postulate that the quantum states of the ``new standard model" $(j, m, m')_q$ are restricted by \ref{knotAssociation}.

The Noether charges of the associated simplest classical knots are then
\begin{empheq}{align}
    Q_w \equiv -k_wm \equiv -k_w\frac{w}{2} \label{qwdef} \\
    Q_r \equiv -k_rm' \equiv -k_r\frac{r+1}{2}  \label{qrdef}
\end{empheq}
where we set o = 1 for the simplest knots. The $k_w$ and $k_r$ are themselves undetermined constants that determine the writhe and rotation charges. The total Noether charge of the simplest knots satisfying (2.2) and (2.3) is then
\begin{equation}
    Q = -k_w\frac{w}{2}-k_r\frac{r+1}{2}
\end{equation}

The association of classical knots with leptons and quarks is somewhat similar to the first association of classical knots by Bohr with the limiting states of the quantized Hydrogen atom. 

There is in addition an empirical relation between the isotopic spin $(t, t_3, t_0)$ of elementary fermions and the knot characterization of the trefoils $(N, w, r)$, as shown in Table 1.

\renewcommand{\arraystretch}{2}

\begin{figure}[H]
\begin{tabular}{r c r r r | p{14mm} c r r c | r}
\multicolumn{10}{c}{\textbf{Table 1:} Empirical Support for $6(t,-t_3,-t_0) = (N,w,r+1)$} \\ 
\hline \hline
 & Elementary Fermions & $t$ & $t_3$ & $t_0$ & Classical Trefoil & $N$ & $w$ & $r$ & $r+1$ & $D^{N/2}_{\frac{w}{2}\frac{r+1}{2}}$ \\[0.2cm]
\hline
\multirow{2}{*}{\hspace{-4pt}leptons $\Bigg \{$} & $(e, \mu, \tau)_L$ & $\frac{1}{2}$ & $-\frac{1}{2}$ & $-\frac{1}{2}$ & & 3 & 3 & 2 & 3  & $D^{3/2}_{\frac{3}{2}\frac{3}{2}}$\\
& $(\nu_e, \nu_{\mu}, \nu_{\tau})_L$ & $\frac{1}{2}$ & $\frac{1}{2}$ & $-\frac{1}{2}$ & & 3 & $-3$ & 2 & 3 & $D^{3/2}_{-\frac{3}{2}\frac{3}{2}}$\\
\multirow{2}{*}{quarks $\Bigg \{$} & $(d, s, b)_L$ & $\frac{1}{2}$ & $-\frac{1}{2}$ & $\frac{1}{6}$ & & 3 & 3 & $-2$ & $-1$ & $D^{3/2}_{\frac{3}{2}-\frac{1}{2}}$\\
& $(\Bar{u}, \Bar{c}, \Bar{t})_L$ & $\frac{1}{2}$ & $\frac{1}{2}$ & $\frac{1}{6}$ & & 3 & $-3$ & $-2$ & $-1$ & $D^{3/2}_{-\frac{3}{2}-\frac{1}{2}}$\\ [0.2cm]
\hline
\end{tabular}

\caption*{{The symbols $(\quad)_L$ designate the left chiral states in the usual notation. The topological labels $(N,w,r)$ on the right side of the table provide a natural way to label the same chiral states. Note that $D^{3/2}_{\frac{w}{2}\frac{r+1}{2}}$, which labels the chiral states, correlates with the elementary fermions on the left.}}
\end{figure}

The chiral states $(N, w, r+1)$ which may be described as 2d projections of 3d trefoils, while the corresponding $(t, -t_3, -t_0)$ states, which may be described as 2d projections of quark states of spin, together suggest the possible relevance of an early two dimensional space (for which there may be independent evidence\cite{carlip}).

There is also the empirical relation \ref{empirical}, which is shown in Tables 1 and 2.

\begin{equation}
    (j, m, m')_q = 3(t, -t_3, -t_0) \label{empirical}
\end{equation}

Only for the particular row-to-row correspondences shown in Table 1 does \ref{empirical} hold, i.e., \emph{each of the four families of fermions labelled by $(t_3, t_0)$ is uniquely correlated with a specific $(w,r)$ classical knot, and therefore with a specific state $D^{3/2}_{\frac{w}{2}\frac{r+1}{2}}(q)$ of the quantum knot.}

Note in Table 1 that the $t_3$ doublets of the standard model now correspond to the writhe doublets ($w = \pm 3$). Note also that with this same correspondence the leptons and quarks form a knot rotation doublet ($r = \pm 2$); the lepton-quark relation depends on $(w,r)$.

Retaining the row to row correspondence described in the Tables, it is then possible to compare in Table 2 the electroweak charges, $Q_e$, \emph{of the most elementary fermions with the total Noether charges, $Q_w + Q_r$, of the simplest quantum knots, which are the quantum trefoils.}

\begin{center}
\begin{tabular} {c r r r c | c l c c c}
\multicolumn{10}{c}{{\textbf{Table 2:} Electric Charges of Leptons, Quarks, and Quantum Trefoils}} \\
\hline \hline
\multicolumn{5}{c |}{{Standard Model}} & \multicolumn{5}{c}{{Quantum Trefoil Model}} \\
\hline
{$(f_1, f_2, f_3)$} & {$t$} & {$t_3$} &{$t_0$} & {$Q_e$} & {$(N,w,r)$} & {$D^{N/2}_{\frac{w}{2}\frac{r+1}{2}}$} & {$Q_w$} & {$Q_r$} & {$Q_w + Q_r$} \\ [0.2cm]
\hline
$(e, \mu, \tau)_L$ & $\frac{1}{2}$ & $-\frac{1}{2}$ & $-\frac{1}{2}$ & $-e$ & $(3,3,2)$ & $D^{3/2}_{\frac{3}{2} \frac{3}{2}}$ & $-k_w \left( \frac{3}{2} \right)$ & $-k_r \left( \frac{3}{2} \right)$ & $-\frac{3}{2}(k_r + k_w)$ \\
$(\nu_e, \nu_{\mu}, \nu_{\tau})_L \hspace{-10pt}$ & $\frac{1}{2}$ & $\frac{1}{2}$ & $-\frac{1}{2}$ & $0$ & $(3,-3,2)$ & $D^{3/2}_{-\frac{3}{2} \frac{3}{2}}$ & $-k_w \left( -\frac{3}{2} \right)$ & $-k_r \left( \frac{3}{2} \right)$ & $\frac{3}{2}(k_w - k_r) $ \\
$(d,s,b)_L$ & $\frac{1}{2}$ & $-\frac{1}{2}$ & $\frac{1}{6}$ & $-\frac{1}{3}e$ & $(3,3,-2)$ & $D^{3/2}_{\frac{3}{2} -\frac{1}{2}}$ & $-k_w \left( \frac{3}{2} \right)$ & $-k_r \left( -\frac{1}{2} \right)$ & $\frac{1}{2}(k_r - 3k_w)$ \\
$(u,c,t)_L$ & $\frac{1}{2}$ & $\frac{1}{2}$ & $\frac{1}{6}$ & $\frac{2}{3}e$ & $(3,-3, -2)$ & $D^{3/2}_{-\frac{3}{2} -\frac{1}{2}} \hspace{-5pt}$ & $-k_w \left( -\frac{3}{2} \right)$ & $-k_r \left( -\frac{1}{2} \right)$ & $\frac{1}{2}(k_r + 3k_w)$ \\
& & \multicolumn{3}{c |}{$ \hspace{-8pt} Q_e = e(t_3+t_0) \hspace{-10pt}$} & \multicolumn{2}{c}{\normalsize $(j, m, m') = \frac{1}{2}(N, w, r + 1)$} & $Q_w = -k_w \frac{w}{2} \hspace{-5pt}$ & $Q_r = -k_r \frac{r+1}{2} \hspace{-15pt}$ & \\
\hline
\end{tabular}
\end{center}

\emph{One sees that \boxed{Q_w + Q_r = Q_e} is satisfied for charged leptons, neutrinos and for both up and down quarks with only a single value of k as follows:}
\begin{equation}
\boxed{k_r = k_w (= k)=\frac{e}{3}} \label{3k}
\end{equation}
and also that $t_3$ isospin and $t_0$ hypercharge then respectively measure the writhe and rotation charges of the associated classical knot:
\begin{equation}
Q_w = et_3 \label{qwet3}
\end{equation}
\begin{equation}
Q_r = e t_0 \label{qret0}
\end{equation}
Then $Q_w + Q_r = Q_e$ becomes by \eqref{qwet3} and \eqref{qret0} an alternative statement of
\begin{equation}
Q_e = e(t_3 + t_0)
\end{equation}
of the standard model. It is important to note that this classification of fundamental particles in Tables 1 and 2 is made by \emph{charge and not by mass}

In SLq(2) measure $Q_e = Q_w + Q_r$ is by \eqref{qwdef} and \eqref{qrdef}:
\begin{equation}
\boxed{Q_e = -\frac{e}{3}(m+m'), \label{qeem}}
\end{equation}
or by (2.3)
\begin{equation}
Q_e = - \frac{e}{6}(w+r+1). \label{qeewr}
\end{equation}
for the quantum trefoils, that represent the elementary fermions.

We can denote the fundamental two dimensional representation of $(j, m, m')_q$ by

\begin{equation}
D^{\frac{1}{2}}_{mm'}(q) = \begin{pmatrix} a & b \\ c & d \end{pmatrix} \label{jhalfrep}
\end{equation}

In the physical model with the \eqref{jhalfrep} coupling we interpret $(a,b,c,d)$ as creation operators for $(a,b,c,d)$ particles, which we term preons. Then we assume that $D^j_{mm'}(a,b,c,d)$ is the creation operator for the state representing the superposition of $(n_a, n_b, n_c, n_d)$ preons. Since we shall regard the preons as fermions, they will also carry an anti-symmetrizing index to satisfy the Pauli principle. 

Here we may additionally generalize by describing some aspects of a generic field theory where the field quanta have two couplings that may be expressed in the coupling matrix
\begin{equation}
\boxed{
\varepsilon_q = \begin{pmatrix}
0 & \alpha_2 \\
-\alpha_1 & 0
\end{pmatrix}. \cite{FinkSchwing}} \label{couplingmatrix}
\end{equation}
where the couplings $\alpha_1$ and $\alpha_2$ are assumed to be dimensionless and real and may be written as
\begin{equation}
(\alpha_1, \alpha_2) \text{ or } (\alpha_2, \alpha_1) = \left( \frac{e}{\sqrt{\hbar c}}, \frac{g}{\sqrt{\hbar c}} \right) \label{couplingconstants}
\end{equation}
where $e$ and $g$ refer to a specific two charge model and have dimensions of an electric charge. We assume that e and g may be energy dependent and normalized at relevant energies. The reference charge is the universal constant $\sqrt{\hbar c}$. 

The fundamental assumption that we make on this coupling matrix is that it is invariant under SLq(2) as follows
\begin{equation}
T \varepsilon_q T^t = T^t \varepsilon_q T = \varepsilon_q \label{slq2invariant}
\end{equation}
where $t$ means transpose and $T$ is a two dimensional representation of SLq(2):
\begin{equation}
T = \begin{pmatrix}
a & b \\
c & d
\end{pmatrix} . \label{2drepslq2}
\end{equation}
where the elements of $T$ obey the knot algebra:
\begin{equation}
\begin{split}
ab = qba \qquad bd = qdb \qquad ad-qbc = 1 \qquad bc &= cb \\
ac = qca \qquad cd = qdc \qquad da -q_1cb = 1 \qquad q_1 &\equiv q^{-1}
\end{split}
\tag{A}
\end{equation}
and
\begin{equation}
\boxed{ q = \frac{\alpha_1}{\alpha_2}}
\end{equation}
so that the two physical couplings fix the algebra through their ratio.

If also
\begin{equation}
\text{det} \hspace{2pt} \varepsilon_q = 1 \label{couplingdet}
\end{equation}
one has
\begin{equation}
\alpha_1 \alpha_2 = 1
\end{equation}
If the two couplings $(\alpha_1, \alpha_2)$ are given by \eqref{couplingconstants}, where $e$ and $g$ are the electroweak and ``gluon''-like couplings, or electric and magnetic couplings respectively, then
\begin{equation}
\boxed{eg = \hbar c} \label{quantize}
\end{equation}

Since $g$ represents magnetic charge, \eqref{quantize} copies the Dirac proposal according to which the magnetic charge is very much stronger than the electric charge. If the magnetic pole is very much heavier as well, it may be observable only at early and not at current cosmological temperatures or at currently achievable accelerator energies.

We may alternatively assume that q-magnetic poles do exist and shall study the possible extension of q-knot symmetry to magnetic charges, in particular as sources of the gravitational field, with \emph{strength} q.
\begin{equation}
    \boxed{g = qe}    
\end{equation}
by (2.17), where q is the deformation parameter of $SL_q(2)$, so that one may assume either (2.20) or (2.21), and $g/e$ may be either $\hbar c / e^2$ or $q$ or both: then $q = \hbar c / e^2$. This new expression for q would directly relate Gamow cosmology to Bohr quantum mechanics, and would be required to be experimentally realized.

\section{Graphical Representation of Corresponding Classical Structures}
The representation of the four classical trefoils as composed of three overlapping preon loops is shown in Figure 1. In interpreting Figure 1, note that the two lobes of all the preon loops make opposite contributions to the rotation, $r$, so that the total rotation of each preon loop vanishes. When the three $a$-preons and $c$-preons are combined to form charged leptons and neutrinos, respectively, as suggested by Harari\cite{harari79}, Shupe\cite{Schupe1}, and Raitio\cite{Raitio1} each of the three labelled circuits is counterclockwise and contributes $+1$ to the rotation while the single unlabeled and shared (overlapping) circuit is clockwise and contributes $-1$ to the rotation so that the total $r$ for both charged leptons and neutrinos is $+2$. In this way the leptons, neutrinos, up and down quarks may be considered composite massive particles that are \emph{sources of the gravitational field}, and are here abreviated by a, b, c, d, in accord with the fundamental representation of $SL_q(2)$. \vspace{-0.3em}

\newgeometry{bottom=2.5cm}
%\begin{figure}[!p] %this line was added by me (Matt). make sure it checks out.
%\caption{Graphical Representation of Corresponding Classical Structures \\ Preonic Structure of Elementary Fermions}
\begin{center}
\normalsize
\begin{tabular}{c | c} 
         \multicolumn{2}{c}{} \\ [-0.54cm]
	\multicolumn{2}{c}{\textbf{Figure 1:} Preonic Structure of Elementary Fermions} \\ [-0.49cm]
	\multicolumn{2}{c}{$Q = -\frac{e}{6}(w+r+o)$, and $(j, m, m') = \frac{1}{2} (N, w, r+o)$} \\ [-0.25cm]
	\multicolumn{2}{c}{$D^j_{mm'} = D^{\frac{N}{2}}_{\frac{w}{2} \frac{r+o}{2}}$}\\
	\begin{tabular}{c c c}
		\multicolumn{3}{r}{ \ul{$(w, r, o)$}} \\ [-0.3cm]
		\multicolumn{3}{l}{Charged Leptons, $D^{3/2}_{\frac{3}{2} \frac{3}{2}} \sim a^3$} \\ [0.2cm]
		 \Large{$\varepsilon^{ijk}$} & $\hspace{17pt} a_j$ & \\ [0.4cm]
		 $\hspace{28pt} a_i$ & & $\hspace{-58pt} a_k$ \\ [-2.8cm]
		 \multicolumn{3}{l}{\xygraph{
!{0;/r1.3pc/:}
!{ \xoverh }
[u(1)] [l(0.5)]
!{\color{black} \xbendu }
[l(2)]
!{\vcap[2] =>}
!{\xbendd- \color{black}}
[d(1)] [r(0.5)]
!{\xunderh }
[d(0.5)]
!{ \xbendr}
[u(2)]
!{\hcap[2]=>}
[l(1)]
!{\xbendl- }
[l(1.5)] [d(0.25)]
!{\xbendl[0.5]}
[u(1)] [l(0.5)]
!{\xbendr[0.5] =<}
[u(1.75)] [l(2)]
!{ \xoverv}
[u(1.5)] [l(1)]
!{\color{black} \xbendr-}
[u(1)] [l(1)]
!{ \hcap[-2]=>}
[d(1)]
!{ \xbendl \color{black}}}} \\ [-1.6cm]
\multicolumn{3}{l}{ \hspace{33pt} \textcolor{red}{
\xygraph{
!{0;/r1.3pc/:}
[u(2.25)] [r(3.25)]
!{\xcapv[1] =>}
[l(2.75)] [u(.75)]
!{\xbendr[-1] =>}
[d(1)] [r(1.25)]
!{\xcaph[-1] =>}
}}}
 \\ [-0.7cm]
		 \multicolumn{3}{r}{\hspace{150pt}$(3,2,1)$}
		 
	\end{tabular}
	&
	\begin{tabular}{c c c}
		\multicolumn{3}{r}{\ul{$(w, r, o)$}} \\ [-0.3cm]
		\multicolumn{3}{l}{$a$-preons, $D^{1/2}_{\frac{1}{2} \frac{1}{2}}$} \\ [1.58cm]
		  & &\hspace{-63pt} $a_i$ \\ [-1.3cm]
		 \multicolumn{3}{l}{\xygraph{
!{0;/r1.3pc/:}
!{\xoverv=>}
[u(0.5)] [l(1)]
!{\xbendl}
[u(2)]
!{\hcap[-2]}
!{\xbendr-}
[r(1)]
!{\xbendr}
[u(2)]
!{\hcap[2]}
[l(1)]
!{\xbendl-}}} \\ [-1.2cm]
\multicolumn{3}{l}{\hspace{39pt} \textcolor{red}{\xygraph{
!{0;/r1.3pc/:}
[d(0.75)] [l(0.75)]
!{\xcaph[-1]=>}
}
}} \\ [-0.7cm]
		 \multicolumn{3}{r}{\hspace{150pt}$(1,0,1)$}
		 
	\end{tabular} \\ [-0.1cm] \hline
\begin{tabular}{c c c}
		\multicolumn{3}{l}{Neutrinos, $D^{3/2}_{-\frac{3}{2} \frac{3}{2}} \sim c^3$} \\ [0.2cm]
		 \Large{$\varepsilon^{ijk}$}& $\hspace{17pt} c_j$ & \\ [0.4cm]
		 $\hspace{28pt} c_i$ & & $\hspace{-58pt} c_k$ \\ [-2.8cm]
		 \multicolumn{3}{l}{\xygraph{
!{0;/r1.3pc/:}
!{\xunderh }
[u(1)] [l(0.5)]
!{\color{black} \xbendu}
[l(2)]
!{ \vcap[2]=>}
!{\xbendd- \color{black}}
[d(1)] [r(0.5)]
!{\xoverh}
[d(0.5)]
!{\xbendr}
[u(2)]
!{\hcap[2]=>}
[l(1)]
!{\xbendl-}
[l(1.5)] [d(0.25)]
!{ \xbendl[0.5]}
[u(1)] [l(0.5)]
!{\xbendr[0.5]=<}
[u(1.75)] [l(2)]
!{\xunderv}
[u(1.5)] [l(1)]
!{\color{black} \xbendr-}
[u(1)] [l(1)]
!{\hcap[-2]=>}
[d(1)]
!{\xbendl \color{black}}}} \\ [-1.5cm]
\multicolumn{3}{l}{\textcolor{red} {\hspace{32pt} \xygraph{
!{0;/r1.3pc/:}
[u(2.25)] [r(3.25)]
!{\xcapv[1] =<}
[l(2.75)] [u(.75)]
!{\xbendr[-1] =<}
[d(1)] [r(1.25)]
!{\xcaph[-1] =<}
}}} \\ [-0.7cm]
		 \multicolumn{3}{r}{\hspace{145pt}$(-3,2,1)$}
		 
	\end{tabular}
	&
	\begin{tabular}{c c c}
		\multicolumn{3}{l}{$c$-preons, $D^{1/2}_{-\frac{1}{2} \frac{1}{2}}$} \\ [1.58cm]
		  & &\hspace{-63pt} $c_i$ \\ [-1.3cm]
		 \multicolumn{3}{l}{\xygraph{
!{0;/r1.3pc/:}
!{\xunderv=>}
[u(0.5)] [l(1)]
!{\xbendl}
[u(2)]
!{\hcap[-2]}
!{\xbendr-}
[r(1)]
!{\xbendr}
[u(2)]
!{\hcap[2]}
[l(1)]
!{\xbendl-}}} \\ [-1.2cm]
\multicolumn{3}{l}{\textcolor{red}{\hspace{39pt} \xygraph{
!{0;/r1.3pc/:}
[d(0.75)] [l(0.75)]
!{\xcaph[-1]=<}
}}} \\ [-0.7cm]
		 \multicolumn{3}{r}{\hspace{145pt}$(-1,0,1)$}
		 
	\end{tabular} \\ [-0.1cm] \hline
\begin{tabular}{c c c}
		\multicolumn{3}{l}{$d$-quarks, $D^{3/2}_{\frac{3}{2} -\frac{1}{2}} \sim ab^2$} \\ [0.2cm]
		 & $\hspace{19pt} b$ & \\ [0.4cm]
		 $\hspace{28pt} a_i$ & & $\hspace{-56pt} b$ \\ [-2.8cm]
		 \multicolumn{3}{l}{\xygraph{
!{0;/r1.3pc/:}
!{\xoverh }
[u(1)] [l(0.5)]
!{\color{black} \xbendu}
[l(2)]
!{\vcap[2]=<}
!{\xbendd- \color{black}}
[d(1)] [r(0.5)]
!{ \xoverv}
[u(0.5)] [r(1)]
!{ \xbendr }
[u(2)]
!{\hcap[2]=<}
[l(1)]
!{\xbendl- }
[u(0.5)] [l(3)]
!{\xoverv}
[u(1.5)] [l(1)]
!{\color{black} \xbendr- }
[l(1)] [u(1)]
!{\hcap[-2]=<}
[d(1)]
!{\xbendl  \color{black}}
[r(2)] [u(0.5)]
!{\xcaph- =>}}} \\[-1.6cm]
\multicolumn{3}{l}{\hspace{42pt} \textcolor{red}{\xygraph{
!{0;/r1.3pc/:}
[u(1.75)] [l(1.5)]
!{\xbendd[-1]=<}
[u(0.75)] [r(1)]
!{\xbendl[-1]=<}
[d(1)] [l(2.25)]
!{\xcaph[-1]=>}
}}} \\ [-0.7cm]
		 \multicolumn{3}{r}{\hspace{140pt}$(3,-2,1)$}
		 
	\end{tabular}
	&
	\begin{tabular}{c c c}
		\multicolumn{3}{l}{$b$-preons, $D^{1/2}_{\frac{1}{2} -\frac{1}{2}}$} \\ [1.58cm]
		  & &\hspace{-63pt} $b$ \\ [-1.3cm]
		 \multicolumn{3}{l}{\xygraph{
!{0;/r1.3pc/:}
!{\xoverv=<}
[u(0.5)] [l(1)]
!{\xbendl}
[u(2)]
!{\hcap[-2]}
!{\xbendr-}
[r(1)]
!{\xbendr}
[u(2)]
!{\hcap[2]}
[l(1)]
!{\xbendl-}}} \\ [-1.5cm]
\multicolumn{3}{l}{\hspace{40pt} \textcolor{red}{\xygraph{
!{0;/r1.3pc/:}
[u(0.75)] [l(0.75)]
!{\xcaph[1]=>}
}}} \\ [-0.1cm]
		 \multicolumn{3}{r}{\hspace{145pt}$(1,0,-1)$}
		 
	\end{tabular} \\ [-0.1cm] \hline
\begin{tabular}{c c c}
		\multicolumn{3}{l}{$u$-quarks, $D^{3/2}_{-\frac{3}{2} -\frac{1}{2}} \sim cd^2$} \\ [0.2cm]
		 & $\hspace{19pt} d$ & \\ [0.4cm]
		 $\hspace{28pt} c_i$ & & $\hspace{-56pt} d$ \\ [-2.8cm]
		 \multicolumn{3}{l}{\xygraph{
!{0;/r1.3pc/:}
!{\xunderh }
[u(1)] [l(0.5)]
!{\color{black} \xbendu}
[l(2)]
!{\vcap[2] =<}
!{\xbendd- \color{black}}
[d(1)] [r(0.5)]
!{\xunderv}
[u(0.5)] [r(1)]
!{\xbendr}
[u(2)]
!{\hcap[2] =<}
[l(1)]
!{\xbendl-}
[u(0.5)] [l(3)]
!{\xunderv}
[u(1.5)] [l(1)]
!{\color{black} \xbendr-}
[l(1)] [u(1)]
!{\hcap[-2]=<}
[d(1)]
!{\xbendl \color{black}}
[r(2)] [u(0.5)]
!{\xcaph- =>}}} \\[-1.75cm]
\multicolumn{3}{l}{\hspace{44pt}\textcolor{red}{\xygraph{
!{0;/r1.3pc/:}
[u(1.75)] [l(1.5)]
!{\xbendd[-1]=>}
[u(0.75)] [r(1)]
!{\xbendl[-1]=>}
[d(1)] [l(2.25)]
!{\xcaph[-1]=<}
}}} \\ [-0.7cm]
		 \multicolumn{3}{r}{\hspace{140pt}$(-3,-2,1)$}
		 
	\end{tabular}
	&
	\begin{tabular}{c c c}
		\multicolumn{3}{l}{$d$-preons, $D^{1/2}_{-\frac{1}{2} -\frac{1}{2}}$} \\ [1.58cm]
		  & &\hspace{-63pt} $d$ \\ [-1.3cm]
		 \multicolumn{3}{l}{\xygraph{
!{0;/r1.3pc/:}
!{\xunderv=<}
[u(0.5)] [l(1)]
!{\xbendl}
[u(2)]
!{\hcap[-2]}
!{\xbendr-}
[r(1)]
!{\xbendr}
[u(2)]
!{\hcap[2]}
[l(1)]
!{\xbendl-}}} \\ [-1.5cm]
\multicolumn{3}{l}{\hspace{39pt} \textcolor{red}{\xygraph{
!{0;/r1.3pc/:}
[u(0.75)] [l(0.75)]
!{\xcaph[1]=<}
}}} \\ [-0.1cm]
		 \multicolumn{3}{r}{\hspace{140pt}$(-1,0,-1)$} \\		 
	\end{tabular} \\ [-0.43cm]
	\multicolumn{2}{c}{The clockwise and counterclockwise arrows are given opposite weights $(\mp 1)$ respectively.} \\ [-0.51cm]
	\multicolumn{2}{c}{The (rotation/writhe charge) is measured by the sum of the weighted (black/red) arrows.} \\
	[-0.51cm]
	\multicolumn{2}{c}{The central loops of the trefoils contribute oppositely to the rotation of the complete trefoil.}
	%\multicolumn{2}{c}{All leptons, neutrinos, quarks, and preon states are here represented by $D^j_{m m'}(q)$}
	\end{tabular}
\end{center}
%\end{figure}

\restoregeometry
%\advance\vsize by -2cm % Return old margins and page height
%\advance\voffset by 2cm % Return old margins and page height

\setlength{\baselineskip}{1.6\baselineskip}

For quarks the three labelled loops contribute $-1$ and the shared loop $+1$ so that $r=-2$, as required.

In each case the three preons that form a lepton trefoil contribute their three negative rotation charges. The geometric and charge profile of the lepton trefoil is thus similar to the geometric and charge profile of a triatomic molecule composed of neutral atoms since the valence electronic charges of the atoms, which cancel the nuclear electronic charges of the atoms, are shared among the atoms to create the chemical binding of the molecule just as the negative rotation charges which cancel the positive rotation charges of the preons are shared among the preons to create the preon binding of the trefoils. There is a similar correspondence between quarks and antimolecules.

We next display the general representation $D^j_{m m'}(q)$ of the algebra $SL_q(2)$ in the Weyl monomial basis just as in the current standard model.
\begin{equation}
\boxed{ {D^j_{mm'}(q) = \sum_{n_a, n_b, n_c, n_d} A(q | n_a, n_b, n_c, n_d) a^{n_a} b^{n_b} c^{n_c} d^{n_d} \label{long}}}
\end{equation}

where $a, b, c, d$ satisfy the algebra (A) and the sum  on $n_a, n_b, n_c, n_d$ is over all positive integers and zero that satisfy the following equations: \cite{finkelstein14a}
\begin{empheq}[box=\fbox]{align}
    n_a + n_b + n_c + n_d &= 2j \label{n2j}\\
    n_a + n_b - n_c - n_d &= 2m \label{n2m}\\
    n_a - n_b + n_c - n_d &= 2m' \label{n2m'}
\end{empheq}

and

\begin{equation}
A (q \vert n_a n_b n_c n_d) = \left [ \frac{ \langle n_+ ' \rangle_1 ! \langle n_- ' \rangle_1 ! }{\langle n_+ \rangle_1 ! \langle n_- \rangle_1 !} \right ]^{\frac{1}{2}} \frac{ \langle n_+ \rangle_1 !}{\langle n_a \rangle_1 ! \langle n_b \rangle_1 !} \frac{\langle n_- \rangle_1 !}{\langle n_c \rangle_1 ! \langle n_d \rangle_1 !}.
\end{equation}
where $n_\pm = j \pm m$, $n ' _\pm = j \pm m '$, and $\langle n \rangle _1 = \frac{q_1 ^n - 1}{q_1 ^1 - 1}$ and $q_1 = q^{-1}$

The two dimensional representation, $T$, introduced by \eqref{2drepslq2} now reappears as the $j = \frac{1}{2}$ fundamental representation
\begin{equation}
\boxed{D^{\frac{1}{2}}_{mm'}(q) = \begin{pmatrix} a & b \\ c & d \end{pmatrix} \label{}}
\end{equation}
In the physical model with the \eqref{couplingmatrix} coupling we interpret $(a,b,c,d)$ in \eqref{long} as creation operators for $(a,b,c,d)$ particles, which we have termed preons. Then $D^j_{mm'}(a,b,c,d)$ is the creation operator for the state representing the superposition of $(n_a, n_b, n_c, n_d)$ gravitational preons. Since we shall regard the preons as fermions, they will also carry an anti-symmetrizing index to satisfy the Pauli principle.

\emph{In proceeding to the quantization of the gravitational field we represent its sources as Standard Model x Knot Model sources ($SL_q(2)$) where the particles of the Knot Model are the preons (a, b, c, d) described in Fig. 1 as $D^{3/2}_{m, m'}(q)$ and also represented as twisted loops as the auxiliary knot particles}. The physical particles are here graphically represented by closed loops of energy-momentum with $n \frac{e}{3}$ charge where n is the number of net positive turns of the tangent in one transit of the loop.

\section{Presentation of the Model in the Preon Representation\cite{fink2018}\cite{FinkSchwing}} 

The particles $(a,b,c,d)$ described in the following sections are assumed to be either e or g preons and carry both e and g charges. The knot representation (3.1) of $D^j_{mm'}$ as a function of $(a,b,c,d)$ and $(n_a,n_b,n_c,n_d)$ implies the following constraints on the exponents:

\begin{IEEEeqnarray*}{+rCl+x*}
n_a+n_b+n_c+n_d & = & 2j & (4.1) \\
n_a+n_b - n_c - n_d & =  & 2m & (4.2) \\
n_a-n_b+n_c-n_d & = & 2m' . & (4.3)
\end{IEEEeqnarray*}
The two relations defining the quantum kinematics and giving physical meaning to $D^j_{mm'}$, namely the postulated \eqref{knotAssociation}:
\begin{equation*}
(j,m,m')_q = \frac{1}{2}(N,w, r+o) \quad \text{field (flux loop) description} \tag{4.4}
\end{equation*}
and the semi-empirical \eqref{empirical} shown in the tables.
\begin{equation*}
(j,m,m')_q = 3(t, -t_3, -t_0)_L \quad \text{particle description} \tag{4.5}
\end{equation*}
imply two complementary interpretations of the relations (4.1) -- (4.3). By (4.4) and these equations, one has a \textit{field} description $(N, w, \tilde{r})$ of the quantum state $(j, m, m')_q$ as follows
\begin{equation}
\left.
{\arraycolsep=1.2pt
\begin{array}{rl}
N &= n_a + n_b + n_c + n_d \\
w &= n_a + n_b - n_c - n_d  \\
\tilde{r} \equiv r+o &= n_a - n_b + n_c - n_d
\end{array}
}
\quad \right\} \text{field (flux loop) (N, w, $\tilde{r}$) description} \label{field (flux loop) description}
\tag{4.6}
\end{equation}
In the last line of \eqref{field (flux loop) description}, where $\tilde{r} \equiv r+o$ and $o$ is a parity index, $\tilde{r}$ has been termed ``the quantum rotation," and $o$ the ``zero-point rotation." 

By \eqref{empirical} one has a ``\textit{particle} description" $(t, t_3, t_0)$ of the same quantum state $(j, m, m')$.
\begin{comment}
\begin{align}
t &= \frac{1}{6} (n_a + n_b +n_c +n_d) \\
t_3 &= -\frac{1}{6}(n_a + n_b - n_c - n_d) \quad \text{particle description} \multirow{3}{*}{\}} \\ 
t_0 &= -\frac{1}{6}(n_a - n_b + n_c - n_d)
\end{align}
\end{comment}
%\begin{comment}
\begin{equation}
\left.
{\arraycolsep=1.2pt
\begin{array}{rl}
t &= \frac{1}{6} (n_a + n_b +n_c +n_d)  \\
t_3 &= -\frac{1}{6}(n_a + n_b - n_c - n_d)  \\
t_0 &= -\frac{1}{6}(n_a - n_b + n_c - n_d) 
\end{array}
}
\quad \right\} \text{particle ($t$, $t_3$, $t_0$) description} \label{particle_description}
\tag{4.7}
\end{equation}
\emph{In \eqref{particle_description}, $(t, t_3, t_0)$ are to be read as SLq(2) preon indices agreeing with standard SU(2) $\times$ U(1) notation} only \emph{at $j = \frac{3}{2}$. In general, however, we do not assume $SU(2)\times U(1)$ and instead may assume that $t_3$ measures writhe charge, $t_0$ measures rotation hypercharge and $t$ measures the total preon population or the total number of crossings of the associated classical knot.}

The attempt to invert (4.4) as (4.6) is quite successful but the corresponding attempt to rewrite the semi-empirical (2.5) as $t, t_3, t_0$ is not. We interpret this difference in favor of (4.4) or as a weakness of (4.5).

\section{Interpretation of the Complementary Equations}
There is also an alternative particle interpretation of the flux loop equations \eqref{field (flux loop) description}
\begin{align*}
N = n_a + n_b + n_c + n_d \tag{5.1$N$} \\
w = n_a + n_b - n_c - n_d \tag{5.1$w$} \\
\tilde{r} = n_a - n_b + n_c - n_d \tag{5.1$\tilde{r}$}
\end{align*}

Here the left-hand side with coordinates $(N,w,\tilde{r})$ label a 2d-projected knot, and the right-hand side describes the preon population of the corresponding quantum state. 

\emph{Equation (5.1$N$) states that the number of crossings, $N$, equals the total number of preons, $N'$, as given by the right side of this equation. Since we assume that the preons are fermions, the knot describes a fermion or a boson depending on whether the number of crossings is odd or even.} Viewed as a knot, a fermion becomes a boson when the number of crossings is changed by attaching or removing a geometric curl
\begin{turn}{90}
\xygraph{
    !{0;/r0.75pc/:}
    !{\xunderh}
    [u l(0.75)]!{\xbendu}
    [l (1.5)]!{\vcap[1.5]}
    !{\xbendd-}}
\end{turn}
. This picture is consistent with the view of a geometric curl as an opened preon loop, in turn viewed as a twisted loop
\begin{turn}{90}
\xygraph{
    !{0;/r0.75pc/:}
    !{\xunderh}
    [u l(0.75)]!{\xbendu}
    [l (1.5)]!{\vcap[1.5]}
    !{\xbendd-}
    [l] [d(0.5)]!{\xbendu-}
    [ld]!{\vcap[-1.5]}
    [u] [r(0.5)]!{\xbendd}
}
\end{turn}. Each counterclockwise or clockwise classical curl corresponds to a preon creation operator or antipreon creation operator respectively.
\newline

\subsection{\emph{Gravitational} Preon Numbers}

Since $a$ and $d$ are creation operators for antiparticles with opposite charge and hypercharge, while $b$ and $c$ are neutral antiparticles with opposite values of the hypercharge, we may introduce the \emph{gravitational preon numbers}
\begin{align}
\nu_a &= n_a -n_d \\
\nu_b &= n_b - n_c
\end{align}
Then (5.1$w$) and (5.1$\tilde{r}$) may be rewritten in terms of gravitational preon numbers as
\begin{align}
&\nu_a + \nu_b = w \hspace{3pt} (= -6t_3) \\
&\nu_a - \nu_b = \tilde{r} \hspace{3pt} (=-6t_0)
\end{align}
By (5.3) and (5.4) the conservation of the preon numbers and of the charge and hypercharge is equivalent to the conservation of the writhe and rotation, which are topologically conserved at the 2d-classical level. In this respect, these quantum conservation laws for preon numbers correspond to the classical conservation laws for writhe and rotation.

Eqns. $(5.1N) - (5.1\tilde{r})$may also be interpreted directly in terms of Fig. 2 by describing the right-hand side of these equations as the possible populations of the conjectured preons at these crossings of Fig. 2 and interpreting the left-hand side as parameters of the binding field that links the 3 conjectured preons.
\newline

\section{Summary on the Measure of Charge by SU(2) $\times$ U(1) and by SLq(2)} 
The SU(2)$\times$U(1) measure of charge requires the assumption of fractional charges for the quarks. The SLq(2) measure requires the replacement of the fundamental charge $(e)$ for charged leptons by a new fundamental charge $(e/3)$ or $(g/3)$ for charged preons but then does not require fractional charges for quarks\cite{Schupe1} if quark charge is expressed in terms of the new ``fundamental or minimal charges", i.e. $(e/3)$ or $q(e/3)$.

The SLq(2), or $(j,m,m')_q$ measure, has a direct preon interpretation since $2j$ is the total number of preonic sources, while $2m$ and $2m'$ respectively measure the numbers of writhe and rotation sources of preonic charge.\cite{Fink2} Since $N$, $w$, and $r$ all measure the handedness of the source, charge is measured by the chirality of the source. \emph{The electric charge of the resultant trefoil or of any composite of preons would then be a measure of the chirality generated by the knotting of an original unknotted flux loop of energy-momentum.}

If neutral unknotted flux tubes predated the particles, and the particles were initially formed by the knotting of unknotted flux tubes of energy-momentum, then the simplest fermions that could have formed with topological stability must have had three 2d crossings and therefore three preons, but the topological stability of this trefoil could be protected only as long as spacetime remains 2-dimensional as suggested by the comments on Tables 1 and 2.

The total SLq(2) charge sums the signed two dimensional clockwise and counterclockwise turns that any energy-momentum current makes both at the crossings and in making a single circuit of the 2d-projected knot. This measure of charge, ``knot charge", which is suggested by the leptons and quarks, appears more fundamental than the electroweak isotopic measure that originated in the neutron-proton system, since it reduces the concept of charge to the chirality of the corresponding energy-momentum curve which may also be described as a SLq(2) chirality and in this way reduces charge to a topological concept similar to the way energy-momentum is geometrized by the curvature of spacetime in the Einstein-Hilbert equations or in lattice methods of Regge for solving the differential equations. In this way the energy-momentum conservation may also be formulated graphically.\cite{Regge}

\section{Other Possible Physical Interpretations of Corresponding Quantum States}

\begin{comment}
The point particle $(N', \nu_a, \nu_b)$ representation and the flux loop $(N, w, \tilde{r})$ complementary representation are related by the $\delta$-function transform:
\begin{align}
\tilde{D}^{N'}_{\nu_a, \nu_b}(q, a, b, c, d) = \sum_{N,w,r} \delta(N',N)\delta(\nu_a+\nu_b, w)\delta(\nu_a - \nu_b, \tilde{r}) D^{N/2}_{\frac{w}{2} \frac{\tilde{r}}{2}} (q, a, b, c, d)
\end{align}
\end{comment}

Since one may interpret the elements $(a,b,c,d)$ of the SLq(2) algebra as creation operators for either preonic particles or current loops, the $D^j_{mp}(q)$ may be interpreted as a creation operator for a composite quantum particle composed of either preonic particles $(N', \nu_a, \nu_b)$ or current loops $(N, w, \tilde{r})$ . \emph{These two complementary views of the same particle may be reconciled as describing $N$-preon systems bound by a knotted field having $N$-crossings with the preons at the crossings as illustrated in Figure 2 for $N=3$.} In the limit where the three outside lobes become small or infinitesimal compared to the central circuit, the resultant structure will resemble a three particle system tied together by a string.
\begin{figure}[!tb]
\begin{center}
\normalsize
\begin{tabular}{c | c}
	\multicolumn{2}{c}{\textbf{Figure 2:} Leptons and Quarks Pictured as Three Preons Bound by a Trefoil Field} \\ [-0.0cm]
	\begin{tabular}{c c c}
		\multicolumn{3}{r}{ \ul{$(w, r, o)$}} \\ [-0.3cm]
		\multicolumn{3}{l}{Neutrinos, $D^{3/2}_{-\frac{3}{2} \frac{3}{2}} \sim c^3$} \\ [0.2cm]
		 & $\hspace{17pt} c_j$ & \\ [-.22 cm]
		 \multicolumn{3}{c}{\hspace{-1.655 cm}\vspace{0.22 cm}\textcolor{blue} {\scalebox{2}{\Huge .}}} \\ [-0.6cm]
		 $\hspace{28pt} c_i$ {\hspace{.33cm}\textcolor{blue} {\scalebox{2}{\Huge .}}} \hspace{-1cm} & & {\hspace{-2.65cm}\textcolor{blue} {\scalebox{2}{\Huge .}}} \hspace{2.15cm} $\hspace{-56pt} c_k$ \\ [-2.8cm]
		 \multicolumn{3}{l}{\xygraph{
!{0;/r1.3pc/:}
!{\xunderh }
[u(1)] [l(0.5)]
!{\xbendu}
[l(2)]
!{\vcap[2]=>}
!{\xbendd-}
[d(1)] [r(0.5)]
!{\xoverh}
[d(0.5)]
!{\xbendr}
[u(2)]
!{\hcap[2]=>}
[l(1)]
!{\xbendl-}
[l(1.5)] [d(0.25)]
!{\xbendl[0.5]}
[u(1)] [l(0.5)]
!{\xbendr[0.5]=<}
[u(1.75)] [l(2)]
!{\xunderv}
[u(1.5)] [l(1)]
!{\xbendr-}
[u(1)] [l(1)]
!{\hcap[-2]=>}
[d(1)]
!{\xbendl}}} \\[-1.5cm]
\multicolumn{3}{l}{\textcolor{red} {\hspace{32pt} \xygraph{
!{0;/r1.3pc/:}
[u(2.25)] [r(3.25)]
!{\xcapv[1] =<}
[l(2.75)] [u(.75)]
!{\xbendr[-1] =<}
[d(1)] [r(1.25)]
!{\xcaph[-1] =<}
}}} \\ [-0.7cm]
		 \multicolumn{3}{r}{\hspace{145pt}$(-3,2,1)$}
		 
	\end{tabular}
&
\begin{tabular}{c c c}
		\multicolumn{3}{r}{ \ul{$(w, r, o)$}} \\ [-0.3cm]
		\multicolumn{3}{l}{Charged Leptons, $D^{3/2}_{\frac{3}{2} \frac{3}{2}} \sim a^3$} \\ [0.2cm]
		 & $\hspace{17pt} a_j$ & \\ [-.22 cm]
		 \multicolumn{3}{c}{\hspace{-1.58 cm}\vspace{0.22 cm}\textcolor{blue} {\scalebox{2}{\Huge .}}} \\ [-0.6cm]
		 $\hspace{28pt} a_i$ {\hspace{.33cm}\textcolor{blue} {\scalebox{2}{\Huge .}}} \hspace{-1cm}  & & {\hspace{-2.65cm}\textcolor{blue} {\scalebox{2}{\Huge .}}} \hspace{2.15cm} $\hspace{-58pt} a_k$ \\ [-2.8cm]
		 \multicolumn{3}{l}{\xygraph{
!{0;/r1.3pc/:}
!{\xoverh }
[u(1)] [l(0.5)]
!{\xbendu}
[l(2)]
!{\vcap[2]=>}
!{\xbendd-}
[d(1)] [r(0.5)]
!{\xunderh}
[d(0.5)]
!{\xbendr }
[u(2)]
!{\hcap[2]=>}
[l(1)]
!{\xbendl-}
[l(1.5)] [d(0.25)]
!{\xbendl[0.5]}
[u(1)] [l(0.5)]
!{\xbendr[0.5]=<}
[u(1.75)] [l(2)]
!{\xoverv}
[u(1.5)] [l(1)]
!{\xbendr-}
[u(1)] [l(1)]
!{\hcap[-2]=>}
[d(1)]
!{\xbendl}}} \\ [-1.6cm]
\multicolumn{3}{l}{\hspace{33pt} \textcolor{red}{
\xygraph{
!{0;/r1.3pc/:}
[u(2.25)] [r(3.25)]
!{\xcapv[1] =>}
[l(2.75)] [u(.75)]
!{\xbendr[-1] =>}
[d(1)] [r(1.25)]
!{\xcaph[-1] =>}
}}} \\ [-0.7cm]
		 \multicolumn{3}{r}{\hspace{150pt}$(3,2,1)$}
		 
	\end{tabular} \\ [-0.1cm] \hline
	\begin{tabular}{c c c}
		\multicolumn{3}{l}{$d$-quarks, $D^{3/2}_{\frac{3}{2} -\frac{1}{2}} \sim ab^2$} \\ [0.2cm]
		 & $\hspace{19pt} b$ & \\ [-.22 cm]
		  \multicolumn{3}{c}{\hspace{-1.52 cm}\vspace{0.22 cm}\textcolor{blue} {\scalebox{2}{\Huge .}}} \\ [-0.6cm]
		 $\hspace{28pt} a_i$ {\hspace{.33cm}\textcolor{blue} {\scalebox{2}{\Huge .}}} \hspace{-1cm}  & & {\hspace{-2.65cm}\textcolor{blue} {\scalebox{2}{\Huge .}}} \hspace{2.15cm} $\hspace{-56pt} b$ \\ [-2.8cm]
		 \multicolumn{3}{l}{\xygraph{
!{0;/r1.3pc/:}
!{\xoverh }
[u(1)] [l(0.5)]
!{\xbendu}
[l(2)]
!{\vcap[2]=<}
!{\xbendd-}
[d(1)] [r(0.5)]
!{\xoverv}
[u(0.5)] [r(1)]
!{\xbendr}
[u(2)]
!{\hcap[2] =<}
[l(1)]
!{\xbendl-}
[u(0.5)] [l(3)]
!{\xoverv}
[u(1.5)] [l(1)]
!{\xbendr-}
[l(1)] [u(1)]
!{\hcap[-2]=<}
[d(1)]
!{\xbendl}
[r(2)] [u(0.5)]
!{\xcaph- =>}}} \\ [-1.6cm]
\multicolumn{3}{l}{\hspace{42pt} \textcolor{red}{\xygraph{
!{0;/r1.3pc/:}
[u(1.75)] [l(1.5)]
!{\xbendd[-1]=<}
[u(0.75)] [r(1)]
!{\xbendl[-1]=<}
[d(1)] [l(2.25)]
!{\xcaph[-1]=>}
}}} \\ [-0.7cm]
		 \multicolumn{3}{r}{\hspace{140pt}$(3,-2,1)$}
		 
	\end{tabular}
	&
\begin{tabular}{c c c}
		\multicolumn{3}{l}{$u$-quarks, $D^{3/2}_{-\frac{3}{2} -\frac{1}{2}} \sim cd^2$} \\ [0.2cm]
		 & $\hspace{19pt} d$ & \\ [-.22 cm]
		  \multicolumn{3}{c}{\hspace{-1.75 cm}\vspace{0.22 cm}\textcolor{blue} {\scalebox{2}{\Huge .}}} \\ [-0.6cm]
		 $\hspace{28pt} c_i$ {\hspace{.33cm}\textcolor{blue} {\scalebox{2}{\Huge .}}} \hspace{-1cm}  & & {\hspace{-2.9cm}\textcolor{blue} {\scalebox{2}{\Huge .}}} \hspace{2.15cm} $\hspace{-56pt} d$ \\ [-2.8cm]
		 \multicolumn{3}{l}{\xygraph{
!{0;/r1.3pc/:}
!{\xunderh }
[u(1)] [l(0.5)]
!{\xbendu}
[l(2)]
!{\vcap[2]=<}
!{\xbendd-}
[d(1)] [r(0.5)]
!{\xunderv}
[u(0.5)] [r(1)]
!{\xbendr}
[u(2)]
!{\hcap[2]=<}
[l(1)]
!{\xbendl-}
[u(0.5)] [l(3)]
!{\xunderv}
[u(1.5)] [l(1)]
!{\xbendr-}
[l(1)] [u(1)]
!{\hcap[-2]=<}
[d(1)]
!{\xbendl}
[r(2)] [u(0.5)]
!{\xcaph- =>}}} \\ [-1.75cm]
\multicolumn{3}{l}{\hspace{44pt}\textcolor{red}{\xygraph{
!{0;/r1.3pc/:}
[u(1.75)] [l(1.5)]
!{\xbendd[-1]=>}
[u(0.75)] [r(1)]
!{\xbendl[-1]=>}
[d(1)] [l(2.25)]
!{\xcaph[-1]=<}
}}} \\ [-0.7cm]
		 \multicolumn{3}{r}{\hspace{140pt}$(-3,-2,1)$}
		 
	\end{tabular} \\ [-0.2cm]
\multicolumn{2}{c}{The preons conjectured to be present at the crossings are suggested by the blue dots at the crossings} \\ [-0.4cm]
\multicolumn{2}{c}{ of the lepton-quark diagrams, or at the crossings of any diagram with more crossings.}
\end{tabular}
\end{center}
\end{figure}
\emph{The physical models suggested by Fig. 2 may be further studied with the aid of preon Lagrangians similar to that given in reference 3}. The Hamiltonians of these three body systems may be parametrized by degrees of freedom characterizing both the preons and the binding field that come from the \emph{form factors required by SLq(2) invariance}. The masses of the leptons, quarks, and binding quanta are determined by the eigenvalues of this Hamiltonian in terms of the parameters describing the constituent preons and energy-momentum flux loops. There is currently no experimental guidance at these conjectured energies. These three body systems are, however, familiar in different contexts, namely
\begin{center}
H$^3$ composed of one proton and two neutrons: $PN^2$ \\
$P$ composed of one down and two up quarks: $DU^2$ \\
$N$ composed of one up and two down quarks: $UD^2$ \\
\end{center}
which are similar to
$U$ as $cd^2$ and $D$ as $ab^2$,
where $U$ and $D$ are up and down quarks, presented as three binding preon states. These different realizations of energy-momentum and charge represent different expressions of curvature and chirality, and in particular as displayed by a closed loop of energy-momentum.

\section{Alternate Interpretation}
In the model suggested by Fig. 2 the parameters of the preons and the parameters of the current loops are to be understood as codetermined. On the other hand, in an alternative interpretation of complementarity, the hypothetical preons conjectured to be present in Figure 2 may carry no independent degrees of freedom and may simply describe \emph{concentrations of energy and momentum at the crossings of the energy-momentum tube}. In this interpretation of complementarity, $(t, t_3, t_0)$ and $(N,w, \tilde{r})$ are just two ways of describing the same quantum trefoil of field. \emph{In this picture the preons are bound}, i.e. they do not appear as free particles. This view of the elementary particles as either non-singular lumps of field or as solitons has also been described as a unitary field theory\cite{Raitio1}.

\section{Lower Representations}
We have so far considered the states $j = 3, \frac{3}{2}, \frac{1}{2}$ representing electroweak vectors, leptons and quarks, and preons, respectively. We finally consider the states $j=1$ and $j=0$. Here we shall not examine the higher $j$ states.

In the adjoint representation $j=1$, the particles are the vector bosons by which the $j=\frac{1}{2}$ preons interact and there are two crossings. These vectors are different from the $j=3$ vectors by which the $j=\frac{3}{2}$ leptons and the $j=\frac{3}{2}$ quarks interact.
\bigskip

If $j=0$, the indices of the quantum knot are
\begin{align}
(j,m,m')_q = (0,0,0)
\end{align}
and by the basic rule \eqref{knotAssociation} for interpreting the knot indices on the left chiral fields
\begin{align}
\frac{1}{2}(N,w,\tilde{r}) = (j, m, m')_q &= (0, 0, 0) 
\end{align}
Then the $j=0$ quantum states correspond to classical loops with no crossings $(N=0)$ just as preon states correspond to classical twisted loops with one crossing. Since $N=0$, the $j=0$ states also have no preonic sources of charge and therefore no electroweak interaction.  \emph{It is possible that these }$j=0$ \emph{hypothetical quantum states are realized as (electroweak non-interacting) loops of field flux with} $w=0$, $\tilde{r}= r+o = 0$\emph{, and }$r = \pm1$, $o =\mp1$ \emph{ i.e. with the topological rotation }$r=\pm1$. The two states $(r, o) = (+1, -1)$ and $(-1, +1)$ are to be understood as quantum mechanically coupled.

If, as we are assuming, the leptons and quarks with $j= \frac{3}{2}$ correspond to 2d projections of knots with three crossings, and if the heavier preons with $j= \frac{1}{2}$ correspond to 2d projections of twisted loops with one crossing, then if the $j=0$ states correspond to 2d projections of simple loops with no crossings, one might ask if these particles with no electroweak interactions and which are smaller and heavier than the preons, are among the candidates for ``dark matter." If these $j=0$ particles predate the $j=\frac{1}{2}$ preons, one may refer to them as ``yons" as suggested by the term ``ylem" for primordial matter.

\section{Speculations about an earlier universe and dark matter} \vspace{-0.8em}
One may speculate about an earlier universe before leptons and quarks had appeared, when there was no charge, and when energy and momentum existed only in the SLq(2) $j=0$ neutral state as simple loop currents of gravitational energy-momentum. Then the gravitational attraction would bring some pairs of opposing loops close enough to permit the transition from two $j=0$ loops into two opposing $j=\frac{1}{2}$ twisted loops. A possible geometric scenario for the transformation of two simple loops of current (yons) with opposite rotations into two $j=\frac{1}{2}$ twisted loops of current (preons) is suggested in Fig. 3. Without attempting to formally implement this scenario, one notes according to Fig. 3 that the fusion of two yons may result in a doublet of preons as twisted loops, which might also qualify as Higgs particles.

\newpage

\begin{center}
\normalsize
\begin{tabular}{c  c  c}
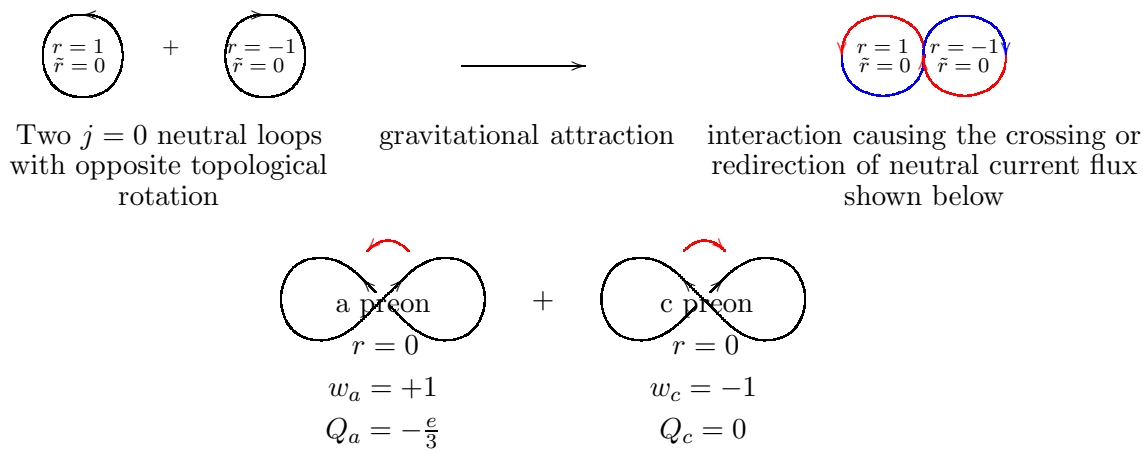
\multicolumn{3}{c}{ \textbf{Figure 3:} Creation of Preons as Twisted Loops} \\
\vspace{-3pt}
\xygraph{
!{0;/r1.3pc/:}
!{\hcap[2]=>}
!{\hcap[-2]}} \hspace{3pt} 
\xygraph{
!{0;/r1.3pc/:}
!{\hcap[2]}
!{ \hcap[-2]=< }} &
\xygraph{
!{0;/r1.3pc/:}
[d(1.25)]
!{\xcaph[3]=<@(0)}} & \hspace{-4pt}
\xygraph{
!{0;/r1.3pc/:}
[d(1)]
!{\color{red} \vcap[2]=> \color{blue}}
!{\vcap[-2]}} \hspace{-7pt} 
\xygraph{
!{0;/r1.3pc/:}
[d(1)]
!{\color{blue} \vcap[2]=<}
!{\color{red} \vcap[-2]=< \color{black}}}\\ [-1.7cm]
\hspace{4pt}\scriptsize{$r=1 \hspace{20pt}+ \hspace{16pt}r=-1$} & & \scriptsize{$\hspace{3pt}r=1 \hspace{8pt} r=-1$} \\ [-0.7cm]
\hspace{0pt}\scriptsize{$\tilde{r}=0 \hspace{32pt} \hspace{16pt}\tilde{r}=0$} & & \scriptsize{$\hspace{0pt}\tilde{r}=0 \hspace{10pt} \tilde{r}=0$} \\
Two $j=0$ neutral loops & gravitational attraction & interaction causing the crossing or  \\ [-0.55cm]
 with opposite topological & & redirection of neutral current flux \\ [-0.55cm]
  rotation & & shown below \\ [-0.55cm]
\end{tabular}
\end{center}
\begin{center}
\begin{tabular}{c c}
\textcolor{red}{\xygraph{
!{0;/r1.3pc/:}
[u(0.75)] [l(0.75)]
!{\xcaph[1]=>}
}} &
\textcolor{red}{\xygraph{
!{0;/r1.3pc/:}
[u(0.75)] [l(0.75)]
!{\xcaph[1]=<}
}} \\ [-1cm]
\xygraph{
!{0;/r1.3pc/:}
!{\xoverv=<}
[u(0.5)] [l(1)]
!{\xbendl}
[u(2)]
!{\hcap[-2]}
!{\xbendr-}
[r(1)]
!{\xbendr}
[u(2)]
!{\hcap[2]}
[l(1)]
!{\xbendl-}} &
\xygraph{
!{0;/r1.3pc/:}
!{\xunderv=<}
[u(0.5)] [l(1)]
!{\xbendl}
[u(2)]
!{\hcap[-2]}
!{\xbendr-}
[r(1)]
!{\xbendr}
[u(2)]
!{\hcap[2]}
[l(1)]
!{\xbendl-}} \\ [-1.1cm]
\multicolumn{2}{c}{+} \\ [-0.95cm]
\hspace{-95pt} & \hspace{-95pt} \\ [-0.9cm]
a preon & \hspace{-1pt} c preon \\ [-0.4cm]
\hspace{2pt}$r=0$ & \hspace{2pt}$r=0$ \\ [-0.4cm]
$w_a = +1$ & $w_c=-1$ \\ [-0.4cm]
$ Q_a = -\frac{e}{3}$ & $ Q_c = 0$  \\
\end{tabular}
\end{center}

In the scenario suggested by Figure 3 the opposing states are quantum mechanically entangled and may undergo gravitational exchange scattering. 

The $\binom{c}{a}$ doublet of Fig. 3 is similar to the Higgs doublet which is independently required by the mass term of the Lagrangian described in reference 3 to be a SLq(2) singlet $(j=0)$ and a SU(2) charge doublet $(t=\frac{1}{2})$. The matrix elements connecting the preon a and c states in Fig. 3 is obviously fundamental for the model. Since the Higgs mass contributes to the inertial mass, one expects a fundamental connection with the gravitational field at this point. 

\begin{comment}
A fraction of the preons $(j= \frac{1}{2})$ produced by the fusion of two yons might in turn combine to form two preon $(j=1)$ and then three preon $(j = \frac{3}{2})$ states with two and three crossings respectively. The $j = \frac{3}{2}$ states would be recognized in the present universe as leptons and quarks. Since, however, the $j = \frac{1}{2}$ and $j=1$ particles with one and two crossings, respectively, are not topologically stable in three dimensions and can relapse into a $j=0$ state with no crossings, the building up process from yons would not produce topologically stable particles before the $j=\frac{3}{2}$ leptons and quarks with three crossings are reached. These remarks apply equally to the e and g sectors.
\end{comment}
 
If at an early cosmological time, only a fraction of the initial gas of quantum loops was converted to preons and these in turn led to a still smaller number of leptons and quarks, then most of the mass and energy of the universe would at the present time still reside in the dark loops while charge and current and visible mass would be confined to structures composed of leptons and quarks. \emph{In making experimental tests for particles of dark matter one might expect the SLq(2) $j=0$ dark loops to be different in mass than the dark neutrino trefoils where $j= \frac{3}{2}$, although both $j=0$ and $j=\frac{3}{2}$ would contribute to the dark matter.}

\bigskip

\section*{Acknowledgements}
I thank E. Abers, C. Cadavid, J. Smit, and S. Mackie for comments.

\end{document}